# Quantized Vibrations in Models of Dibaryons, Pentaquarks and Z* Resonances

David Akers*
*Lockheed Martin Corporation, Dept. 6F2P, Bldg. 660, Mail Zone 6620,
1011 Lockheed Way, Palmdale, CA 93599*
*Email address: David.Akers@lmco.com

The existence of quantized vibrations is utilized to explain narrow nucleon resonances below the pion threshold. A Goldstone boson with a mass of 35 MeV is derived from the experimental dibaryon masses. The quantum of 35 MeV is recognized as part of the Nambu empirical mass series. We present a few explanations to suggest why this light boson is difficult to locate in experiments. The KARMEN timing anomaly is discussed. The vibration-rotational model of elementary particles is reviewed in regards to the Diakonov-Petrov argument against mixing of these modes. The existence of a $\theta(1640)D_{03}$ partner to the $\theta^+(1540)$ is predicted.

## I. INTRODUCTION

In this paper, the existence of quantized vibrations is derived from the experimental dibaryon masses. Previous research utilized the idea of quantized vibrations to explain the spectra of mesons, baryons and exotic baryons. The existence of light quanta was proposed earlier in several models, inclusive of the constituent quark (CQ) model of Mac Gregor and the Goldstone boson model. The vibration-rotational models explain numerous experimental data. In the present work, we suggest that quantized vibrations may account for excited narrow nucleons, the equidistant mass spacings of dibaryons, and the KARMEN timing anomaly. These ideas are discussed in Section II. From the symmetry of Z* resonances among the spectrum of exotic baryons (pentaquarks) and from the existence of quantized vibrations, we are able to predict the existence of a θ(1640)$D_{03}$ partner to the $θ^+$(1540). The analysis for the pentaquarks is presented in Section III. Concluding remarks are made in Section IV.

## II. WALCHER PARTICLE-HOLE MODEL

In recent years, experiments reported narrow excited states of the nucleon below the π threshold [1-3]. Peaks were observed at 1004, 1044, and 1094 MeV with a width of about 5 MeV [1]. The first experiments involved pp → p $π^+$ N*. The second experiments found three peaks at 966, 985, and 1003 MeV [2]. These experiments involved the reaction pd → pp$X_1$, where $X_1$ = N* → γ N, and measured the missing mass $X_1$. Other experiments of electron scattering on hydrogen/deuterium targets have indicated a null signal [4]. L'vov and Workman have argued that these narrow nucleon resonances are not observed in real Compton scattering experiments and must be excluded [5]. Thus, the existence of these reported narrow states has been met with disbelief.

It is noted that Walcher is a member of the collaboration [4], which has indicated no evidence for these narrow states. Earlier, Walcher had proposed a particle-hole model from nuclear physics to explain these states [6]. In this section, we review the particle-hole model and show that the feature of quantized excitations or vibrations is reasonable. Walcher proposed a light "pion" to account for equidistance between masses from all known experiments in Ref. [1-3]. An average mass difference of δm = 21.2 ± 2.6 MeV was calculated as a fit to the mass series 940, 966, 985, 1003, $m_4$, 1044, $m_6$, and 1094 MeV, where $m_4$ = 1023 and $m_6$ = 1069 MeV. Thus, a light quantum of excitation, $m_{light\,π}$ = $2m_q$ = 21 MeV, was proposed by Walcher [6].

It is also noted that the existence of light quanta was proposed earlier in several models, albeit with masses 35 and 70 MeV [7-9]. Furthermore, it is noted that none of the isospins, spins and parities of the reported nucleon states have been measured because of the difficulties with weak signals and measuring angular distributions. Hence, reaching an inference for equidistance between these nucleon masses must be met with caution since there may be different quantum numbers among the states.

From a study of the experimental dibaryon masses, Walcher determined that the dibaryon masses were equidistant and that there existed an excitation quantum of 36.7 MeV as calculated from his fit to the spectrum. There was no attempt to correct for the effects of the binding energies of the dibaryons. However, Mac Gregor has shown that typical binding energies are about 3-4% in many baryons [8]. Therefore, the equidistant hypothesis for the dibaryons should be nearly exact, as shown in Fig. 3 of Ref. [6]. We note also that Walcher calculates incorrectly the intercept and slope in Fig. 3 of Ref. [6].

In Fig. 1 of the present work, we plot the dibaryon masses (in MeV) as a function of the number n = 0, 1, 2, … of the excitation quantum. An intercept is calculated at about 1830 MeV, and a quantum is derived to be 35 MeV for the line in Fig. 1. The dibaryon masses fit a straight line to the formula:

$$m = 1830 \text{ MeV} + n \cdot 35 \text{ MeV}, \tag{1}$$

where n = 0, 1, 2, … It is recognized immediately that the 35 MeV quantum is part of the Nambu empirical mass formula [10]. The 35 MeV quantum, along with the entire Nambu series, was later derived in a modified QCD Lagrangian [9]. The 35 MeV quantum is one-half of the famous 70 MeV quantum proposed by Mac Gregor [7, 8]. The Nambu mass series includes 0, 35, 70, 105, 140, 175, 210, … MeV. The masses of the µ = 105 MeV and π = 140 MeV are easily recognized. Mac Gregor noted m = 70 MeV, B = 140 MeV, F = 210 and X = 420 MeV in his notation [8]. Mac Gregor also had some earlier notation for these quanta, spinors, and cabers [11-13]. We emphasize again that the 35 MeV quantum is in the Nambu empirical mass formula. This quantum was derived in a modified QCD Langrangian, where magnetic monopoles were proposed as Goldstone bosons [9].

For those who may be unfamiliar with the famous 70 MeV quantum of Mac Gregor, it is easily derived from the electron mass and the fine structure constant as:

$$m = m_e/\alpha = 70 \text{ MeV}. \qquad (2)$$

In the constituent quark (CQ) model, Mac Gregor noted that the pion is a composite of two 70 MeV/$c^2$ mass quanta or $m_1 = m_2 = 70$ MeV, where $m = m_1 + m_2 = 140$ MeV. Now that the correct mass for the proposed Goldstone boson is noted as $m_G = 35$ MeV, we turn to the particle-hole model as proposed by Walcher.

We refer the reader to section II in Ref. [6] for the application of the particle-hole model from nuclear physics to the situation of nucleon excitation. Walcher derived the equation of the energy eigenvalues for the constituent boson and current boson. The constituent boson is composed of two constituent quarks, and the current boson is composed of two current quarks. The final equation is in Ref. [6]:

$$1 = \lambda m_G/(E - m_G) + \lambda m_{\pi c}/(E - m_{\pi c}) . \qquad (3)$$

Walcher solves this equation as a function of E and plots the result as Fig. 2 in Ref. [6]. The physical solution appears at $E = m_{\pi c} = 138$ MeV with $\lambda = -0.94$ if one chooses $m_G = 21$ MeV. Walcher states, "Since $\lambda \approx -1$ this solution shows that the creation-annihilation diagram dominates indeed and that the particle-hole interaction is attractive as required" [6]. If one chooses the correct quantum $m_G$ = 35 MeV and E = $m_\pi$ = 140 MeV, the fit is easily obtained with $\lambda$ = -1 as a solution:

$$-1 = m_G/(E - m_G) + m_{\pi c}/(E - m_{\pi c}) \text{, or}$$

$$E^2 = m_G m_{\pi c}, \qquad (4)$$

where we recognize immediately the relativistic energy equation with momentum squared $p(m_G) p(m_{\pi c}) \approx 0$ and with the speed of light c = 1.

If the constituent boson mass $m_G$ is plotted as a function of the current boson mass $m_{\pi c}$ with E = 140 MeV, then we obtained from Eq.(4) the graph with the square markers in Fig. 2. This is the boson-boson curve on the scale of the current quark mass. The current quark mass is one-half of the mass of the current boson mass. A plot of the constituent boson mass $m_G$ is shown as a function of the current quark mass $m_q$; this is the graph with the triangles as markers in Fig. 2. This is the boson-fermion curve on the same scale. Finally, the constituent quark mass is one-half the mass of the constituent boson mass $m_G$. A plot of the constituent quark mass is shown as a function of the current quark mass. This is indicated with solid circles in Fig. 2, and this represents the fermion-fermion curve on the same energy scale.

1) We suggest that the hv = 35 MeV quantum is a radial vibration of the nucleon. Synonymous terms would be a radial oscillation, an excitation, and the monopole ($\lambda$ = 0) or breathing mode [14]. If the nucleon *could* be modeled as a quark or boson condensate, then the 35 MeV quantum *may be* considered a phonon vibration.

2) Walcher discusses eight points about Goldstone bosons, Compton scattering, and lattice physics in section III of Ref. [6]. We shall not review nor comment on the excellent ideas presented in that section. However, we are left with how to explain the appearance of the narrow excited nucleon states and the negative results from electron scattering experiments. If the excited nucleon states are in a monopole (E0) or breathing mode, would we expect photon E1 transitions to dominate at low energies? Moreover, can the monopole (E0) mode be detected in the reactions pp → p $\pi^+$ N* and pd → pp$X_1$? These are questions for the experimentalists to answer.

3) Another pressing question is: why has not the light "pion" (as Walcher uses the term) been seen in experiments? That is, why is the quantum of 35 MeV not freely seen? A few possible explanations are that the quantum is only part of the radial vibration of the excited nucleon or that the quantum is rarely produced in experiments. The first explanation is easily understood as a possibility. The latter explanation deserves further exploration.

4) The second explanation has the rare appearance of a neutral particle in the KARMEN timing anomaly [15]. They found an anomaly in the time distribution of their events which suggested a neutral particle of mass ≈ 33.9 MeV. The explanation for these events was a rare decay of the pion: $\pi^+ \to \mu^+$ X. Although this decay *may be* possible, it would suggest that the neutral particle has a spin ½. Subsequent experiments have found no evidence for this particular decay mode [16]. Atchison *et al.* have suggested a beam-correlated neutron background to explain the timing anomaly [17]. Additional experiments are planned to search for the rare decay $\pi^+ \to \mu^+ Q^0$ at MiniBooNE [18]. If the neutral particle is a quantized excitation of the nucleon, then we would not expect the rare decay of the pion to this neutral particle. In fact, the KARMEN timing anomaly may be due to an excitation of the protons in the beam or of the nucleons in the target.

5) In regards to the second explanation for the rare appearance of the quantum with a mass of 35 MeV in experiments, we propose a model with the photon Compton scattering off two nucleons. In Fig. 3, we show a photon with energy E = 140 MeV incident on a nucleon $N_1$ scattering center. The nucleon $N_1$ scatters through an angle $\phi_1$ relative to the beam direction and a quantum of 70 MeV is scattered at $90^0$. This is the famous right angle scattering studied by Compton. The quantum of 70 MeV is now incident on a nucleon $N_2$ scattering center. The nucleon $N_2$ scatters through an angle ($90^0$ - $\phi_2$) relative to the beam direction and a quantum of 35 MeV is scattered at $90^0$ into the beam direction. The Compton wavelength of the quantum of 70 MeV is 1.77 fm. Thus, the quantum of 35 MeV is lost in the backward or forward beam direction. There is little angular distribution for the quantum of 35 MeV. The nucleon $N_2(90^0 - \phi_2)$ would scatter with a time delay relative to the nucleon $N_1(\phi_1)$, and the time delay would be dependent upon the mean distance between the two scattering centers. Of course, we could have devised a scattering experiment with an initial photon of energy E = 70 MeV, and the first scattered quantum would have a scattered energy of 35 MeV at right angles. The Compton wavelength of the 35 MeV quantum would be 2(1.77) = 3.54 fm. What would be the mean distance that the quantum of 35 MeV could travel before scattering off a second nucleon? The resulting energy of the secondary scattering would be ½(35) = 17.5 MeV, which is very close to the proposed quantum as suggested by Walcher [6]. Thus, the question remains to whether the proposed light quanta can be extracted from the incident beams.

## III. VIBRATION-ROTATIONAL MODELS

The earliest known research into a study of the rotational spectra of baryons and mesons can be attributed to Mac Gregor from the 1970s [12,19]. In the work of Mac Gregor, a constituent-quark (CQ) model was proposed that featured baryons and mesons as rotational states, which suggested a correlation between increasing L values and increasing energies in these resonances. These resonances followed the expected interval rule of L(L + 1) for rotational spectra. Recently, the rotational spectra of mesons and baryons were studied in Ref. [20]. The experimental particle data are derived from the *Review of Particle Properties* [21]. Additionally, the rotational spectra of exotic baryons were also studied [22].

The nature of the quantum with mass 35 MeV is discussed in Ref. [22], and it involves the quantized vibrations of the nucleons. The evidence is shown for nucleon vibrations in steps of 70 and 140 MeV [8, 13, 20]. Many models of baryons utilize a Hamiltonian which involves harmonic oscillator potentials as the confining force for short range distances [23]. The 3-dimensional harmonic oscillator potentials result in quantized vibrations with the energy $E = (N + 3/2)h\nu$ [23]. It is the general belief that the vibrations and rotations will mix in baryon states. This result comes from the molecular model of baryons. In the molecular model, the vibration and rotational modes are mixed in the Hamiltonian calculations. Recently, Williams and Gueye derived a radial Schrödinger equation and the energy eigenvalues for the mixing between these modes [24].

The quark molecular model (QMM) of Williams and Gueye cannot be correct, because their energy eigenvalues, $E_{nl} = (2n + l + 3/2)\omega$, involve the mixing of oscillator vibrations and the rotational modes. The experimental data does not support this mixing (see Figs. 1-4 in Ref. [20]). In fact, Diakonov and Petrov have argued that the mixing is small between the rotational modes and vibrations [25]. They state that "the rotational level spacing is actually irrelevant to the shift of the rotational energy owing to vibrations" [25]. The quantized vibrations are relatively independent of the rotations in baryons. In Fig. 4, we show the rotational spectra of the nucleons from Ref. [20]. We note the linear relationship between the energy and the $L(L + 1)$ interval rule in Fig. 4. If the vibrations mixed with the rotational modes in the Hamiltonian, then these lines would not be parallel (see Figs. 1-4 in Ref. [20]). Thus, the quark molecular model (QMM) of Williams and Gueye does not give the correct relationship between vibrations and rotational modes.

The experimental data, as shown in Ref. [20], supports the conclusion of Diakonov and Petrov. This is *not* a surprising result. In fact, Mac Gregor would not have been able to derive his models for the rotational spectra of mesons and baryons in the 1970s if there is large mixing between the vibrations and rotational modes. Moreover, Nichitiu studied the rotational picture for dibaryons and found linear relationships between the energy and the $L(L + 1)$ interval rule [26]. He noted equidistant mass spacings of 177 MeV between the lines for rotational energies (see Eq.(5) in Ref. [26]). Likewise, Nichitiu found the excitations between $Z_0$ resonances and $\Lambda_r$ resonances to be $m_{Z0} - m_{\Lambda r}$ = 177 MeV. Several sets of data have quantized vibrations $\approx$ 177 MeV in Nichitiu's paper [26]. We recognize that this value of 177 MeV, within experimental uncertainty, is near 140 + 35 = 175 MeV = n · 35 MeV, where n = 5 in the Nambu empirical mass formula. We note that the figure in Ref. [26] should have $L(L + 1)$ as the abscissa and not $L(L - 1)$ as shown.

Thus, Nichitiu found a quantized vibration of approximately 175 MeV between rotational lines.

Additionally, quantized vibrations are suggested in the rotational spectra of exotic baryons with pentaquark configurations [22]. In Ref. [22] the rotational spectra for pentaquarks are suggested with quantized vibrations in units of the 70 MeV. It was also predicted that there would be a $\theta(1610)D_{03}$ partner to the $\theta(1540)$ baryon [22]. A rotational energy of $E_{rot}$ = 18.5 MeV is calculated based upon the $L(L + 1)$ interval rule. The mass for the $D_{03}$ may be low from initial considerations. If the rotational energy is calculated to be $E_{rot}$ = 25 MeV, then the partner would have a mass of 1640 MeV. That is, $\theta(1640)D_{03}$ would be the close partner to the $\theta(1540)$ baryon. In Fig. 5, we show the rotational spectra of the theta baryons mixed with a few Z* resonances. This spectrum is borrowed from Ref. [22]. The $Z^*(1788)D_{03}$ is found in the analyses of Hyslop *et al.* [27]. The $Z^*(1490)D_{03}$ is found in the speed plot analyses of Kelkar *et al.* (see Fig. 1(b) in Ref. [28]). We note the equidistant spacing between these states and the excitation energies in 70 MeV units. The $\theta(1640)D_{03}$ is half-way between the two Z* resonances as shown in Fig. 5. The spacing is on the order of the pion mass $\approx$ 148-150 MeV. Thus, there is a quantized excitation or vibration between these $D_{03}$ states, marking the existence of the $\theta(1640)D_{03}$ partner to the $\theta(1540)$ baryon.

We note also that Jennings and Maltman (JM) also predicted the $\theta(1640)D_{03}$ partner to the $\theta(1540)$ baryon [29]. In a model of the Goldstone boson exchange (GBE) between constituent quarks, JM calculated the mass of this partner and *compared it to the $D_{03}$ resonance at 1788 MeV*. We quote JM: "so the GB prediction gives a prediction for the negative parity excitation energy which is a factor of > 2 too small, though the quantum numbers of the lowest-lying negative parity state are in agreement with the experimental claim" [29]. JM do not provide details of their calculations in Ref. [29], but there are plans to present the calculations soon [30]. JM speculated that the Z* resonance at 1788 MeV and the $\theta(1640)D_{03}$ were the same predicted partner to the $\theta(1540)$ baryon. In fact, the Z* resonance at 1788 MeV is a separate state relative to the $\theta(1640)D_{03}$, as shown in Fig. 5, so that their original calculations are correct for the predicted mass. Finally, we note that Dudek and Close also predicted, with less precision in their calculations, a $D_{03}$ partner with a mass in the range 1540-1680 MeV [31].

## IV. CONCLUSION

In this paper, we presented evidence for the existence of quantized vibrations from the experimental dibaryon masses. Previous research utilized the idea of quantized vibrations to explain the spectra of mesons, baryons and exotic baryons. The existence of light quanta was proposed earlier in the constituent quark (CQ) model of Mac Gregor and the Goldstone boson model, where the magnetic monopole was treated as a boson in a modified QCD Lagrangian. The existence of the 35 MeV quantum was derived from the modified QCD Lagrangian and forms the basis for the Nambu empirical mass formula. The vibration-rotational models explained numerous experimental data in previous studies. In the present work, we suggested that quantized vibrations may account for excited narrow nucleons, the equidistant mass spacings of dibaryons, and the KARMEN timing anomaly. The vibration-rotational modes were shown to be independent and involved no mixing in the rotational spectra of nucleons. From the symmetry of Z* resonances among the spectrum of exotic baryons (pentaquarks) and from the existence of quantized vibrations, we were able to predict the existence of a $\theta(1640)D_{03}$ partner to the $\theta^+(1540)$.

## ACKNOWLEDGEMENT

The author wishes to thank Dr. Malcolm Mac Gregor, retired from the University of California's Lawrence Livermore National Laboratory, for his encouragement to pursue the CQ Model, and he wishes to thank Dr. Paolo Palazzi of CERN for his interest in the work and for e-mail correspondence.

**FIGURE CAPTIONS**

**Fig. 1.** The experimental dibaryon masses against the number of excitation quanta with unit mass 35 MeV. Experimental data are from Fig. 18 in Ref. [32], where the average mass difference is computed to be 34.8 MeV for the quantum (*cf.* Fig. 8 in Ref. [13]).

**Fig. 2.** The constituent quark mass as a function of the current quark mass is indicated with circles. The constituent quark mass as a function of the Goldstone-boson mass is noted by the triangles. The Goldstone-boson mass as a function of the constituent "pion" mass is represented by the square points.

**Fig. 3.** A model for Compton scattering is proposed to explain the difficulty in locating the 35 MeV quantum.

**Fig. 4.** The rotational spectra of the nucleons are shown as a function of the $L(L + 1)$ interval rule.

**Fig. 5.** The spectra of exotic baryons are shown with quantized excitations. The $\theta(1640)D_{03}$ is shown equidistant between the two $Z^*$ resonances.

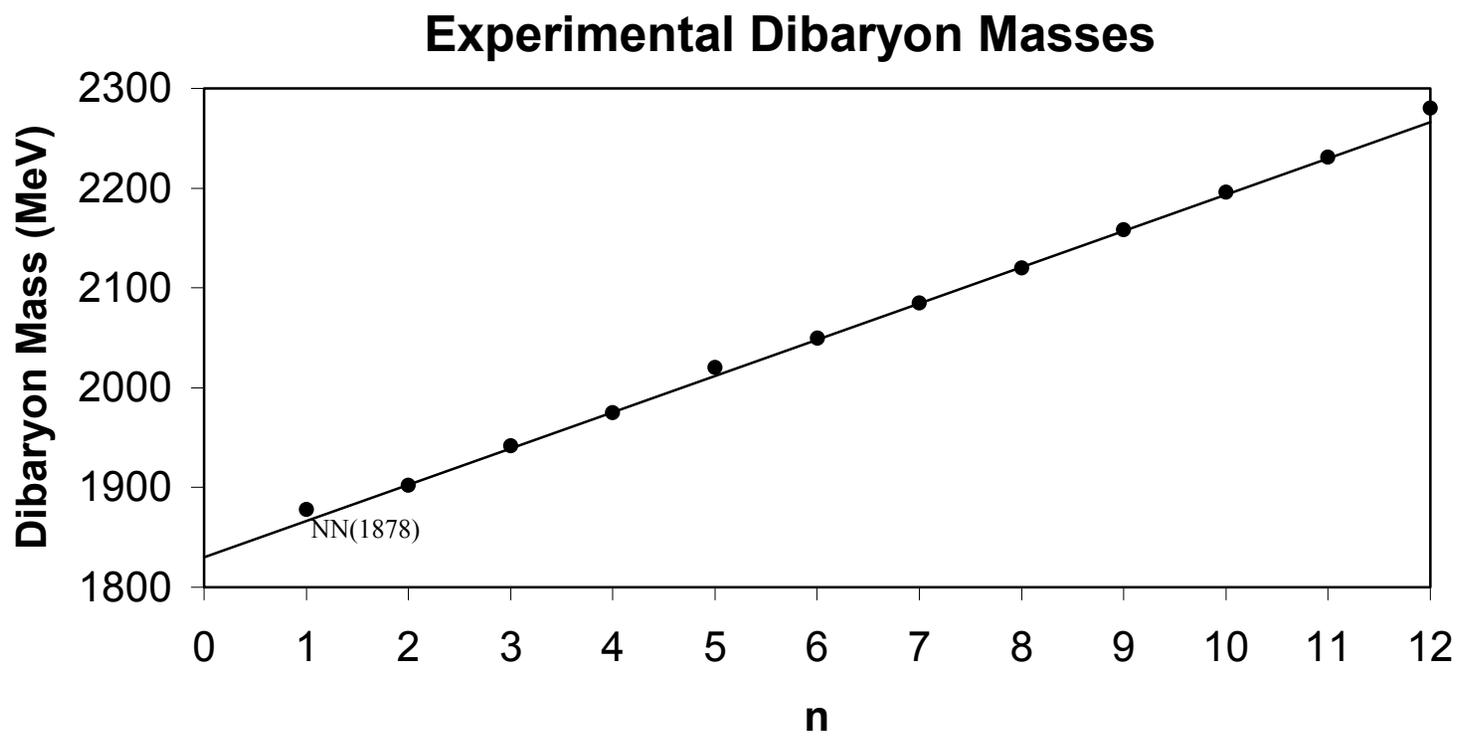

Fig. 1.

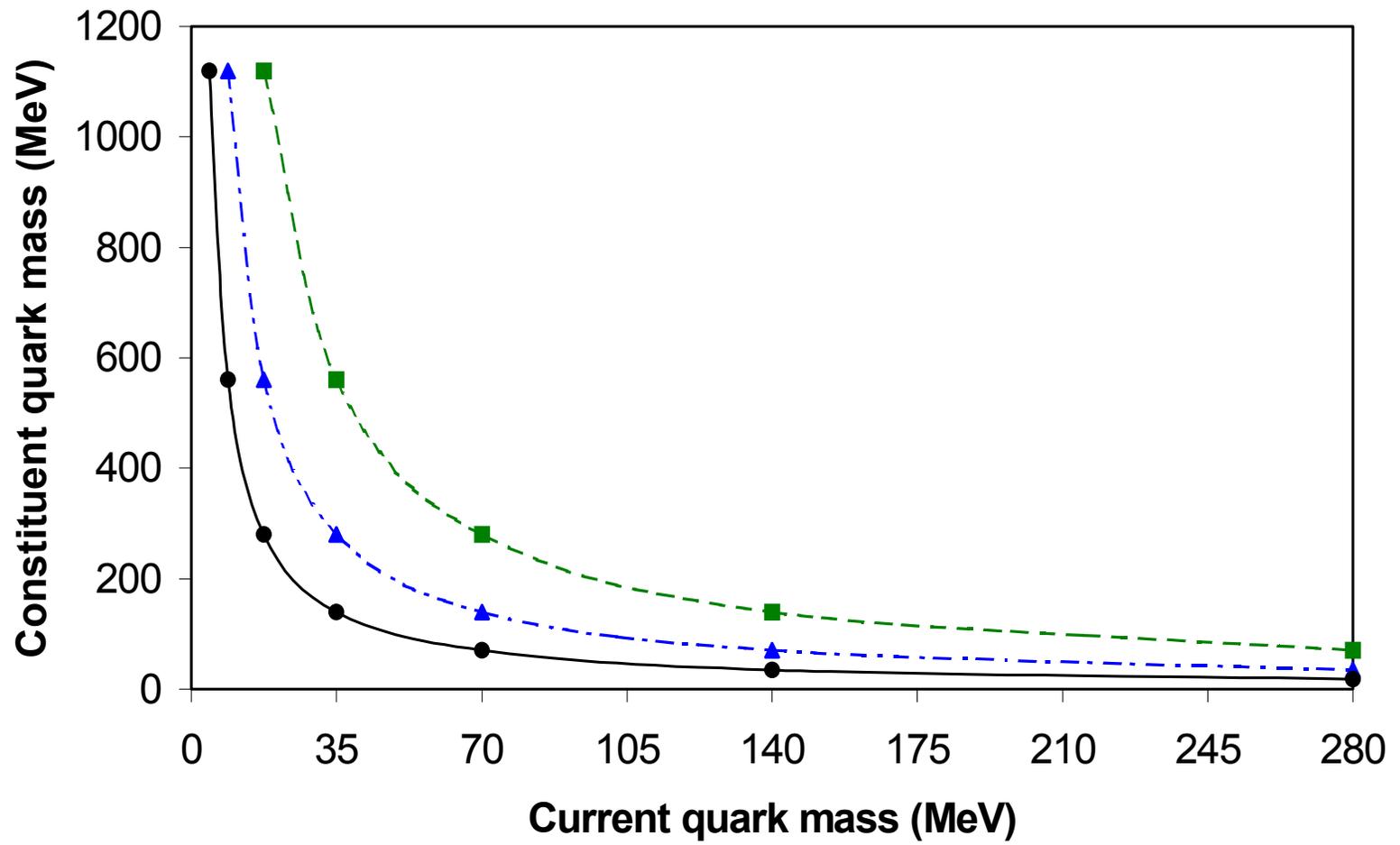

**Fig. 2.**

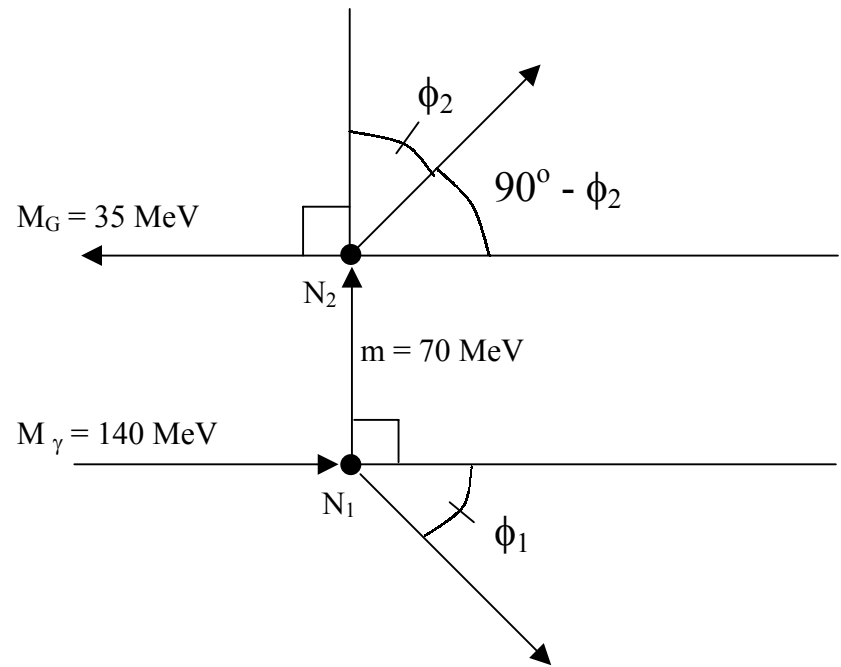

**Fig. 3.**

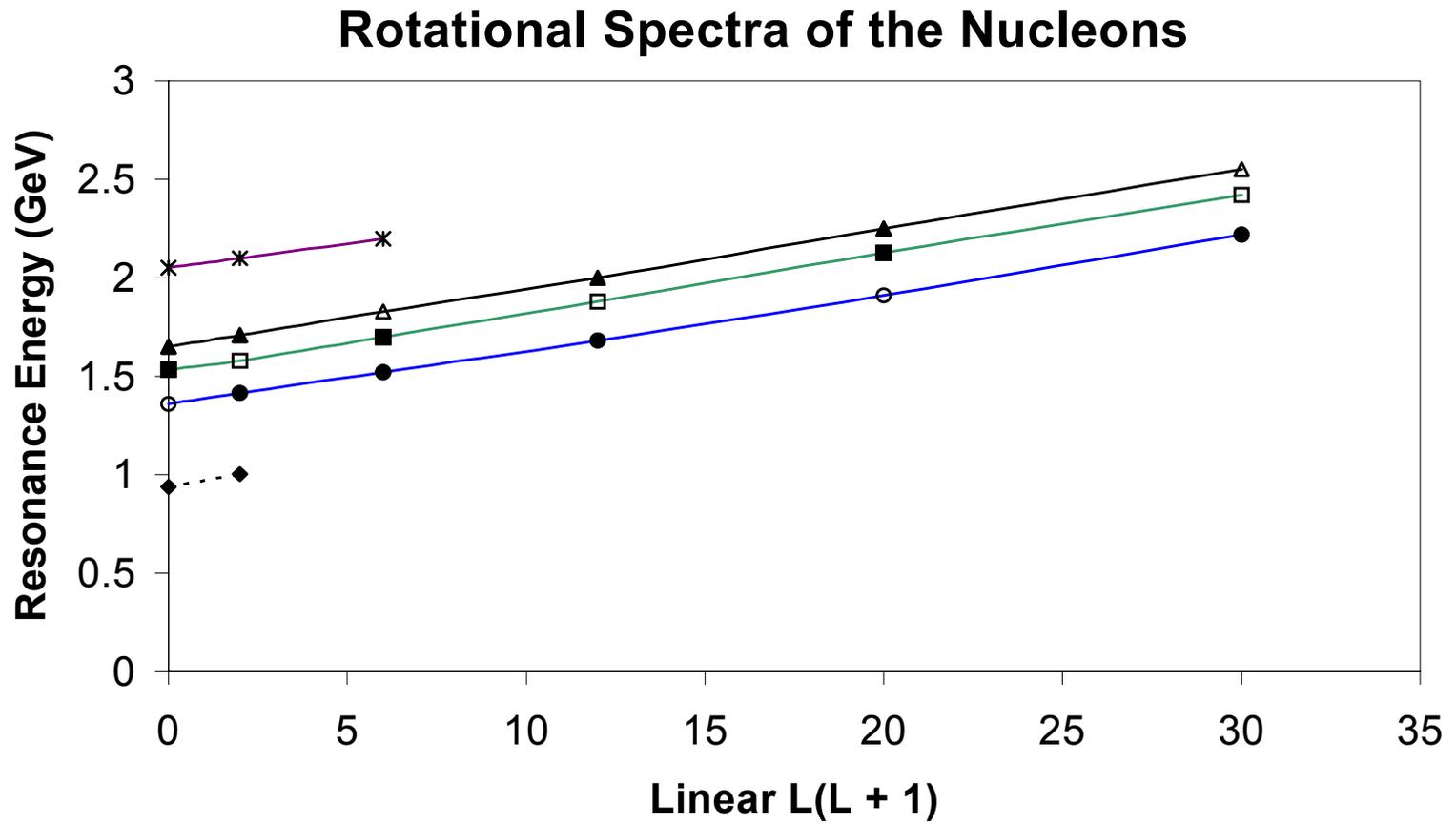

**Fig. 4.**

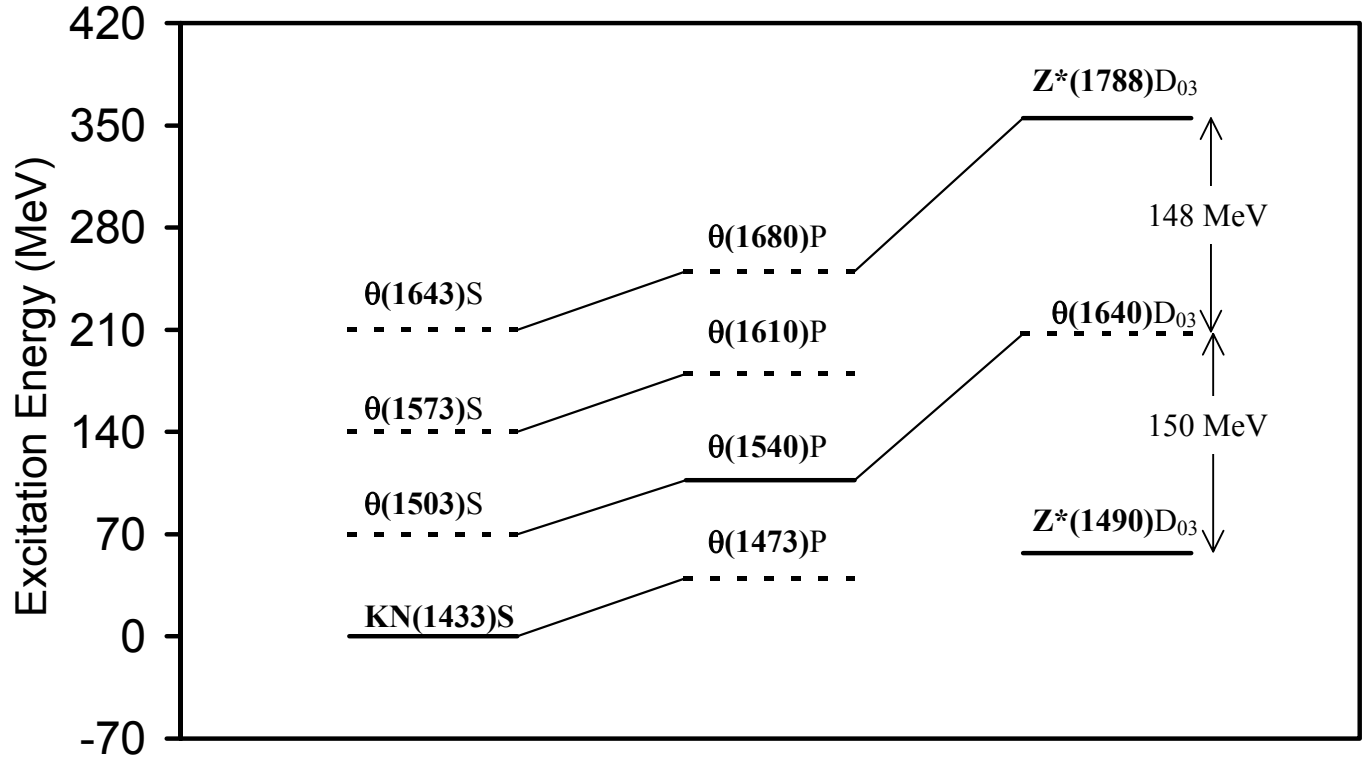

**Fig. 5.**